\documentclass[12pt,preprint]{aastex}
\newcommand{\nraoblurb}{The National Radio Astronomy Observatory is
a facility of the National Science Foundation operated under cooperative
agreement by Associated Universities, Inc.}

\newcommand{\arcmper}{\rlap.{^{\prime}}}

\newcommand{\rgal}{\ensuremath{{R_{\rm gal}}}}

\newcommand{\mhz}{\ensuremath{\,{\rm MHz}}}
\newcommand{\ghz}{\ensuremath{\,{\rm GHz}}}
\newcommand{\kel}{\ensuremath{\,{\rm K}}}
\newcommand{\K}{\ensuremath{\,{\rm K}}}

\newcommand{\m}{\ensuremath{\,{\rm m}}}

\newcommand{\percc}{\ensuremath{\,{\rm cm^{-3}}}}
\newcommand{\kpc}{\ensuremath{\,{\rm kpc}}}
\newcommand{\pc}{\ensuremath{\,{\rm pc}}}
\newcommand{\kms}{\ensuremath{\,{\rm km\, sec^{-1}}}}
\newcommand{\msun}{\ensuremath\,M_\odot}
\newcommand{\s}{\,s}

\newcommand{\jy}{\,Jy}

\newcommand{\kapi}{\ensuremath{\kappa_i}}
\newcommand{\hp}{\ensuremath{{\rm H}^+}}
\newcommand{\he}[1]{\ensuremath{^#1{\rm He}}}
\newcommand{\heo}[1]{\ensuremath{^#1{\rm He}^{\rm o}}}
\newcommand{\hep}[1]{\ensuremath{^#1{\rm He}^+}}
\newcommand{\hepp}[1]{\ensuremath{^#1{\rm He}^{++}}}
\newcommand{\her}[1]{\ensuremath{^#1{\rm He}/{\rm H}}}
\newcommand{\hepr}[1]{\ensuremath{^#1{\rm He}^{+}/{\rm H}^+}}

\newcommand{\hal}{\ensuremath{{\rm H}91\alpha}}

\newcommand{\heal}{\ensuremath{{\rm He}91\alpha}}
\newcommand{\hebet}{\ensuremath{{\rm He}114\beta}}

\newcommand{\halk}{\ensuremath{{\rm H}70\alpha}}
\newcommand{\hbetk}{\ensuremath{{\rm H}88\beta}}
\newcommand{\healk}{\ensuremath{{\rm He}70\alpha}}
\newcommand{\hebetk}{\ensuremath{{\rm He}88\beta}}

\newcommand{\nexpo}[2]{\ensuremath{#1 \times 10^{#2}}}

\newcommand{\hei}{He~{\sc i}}
\newcommand{\hii}{H~{\sc ii}}
\newcommand{\heii}{He~{\sc ii}}
\newcommand{\y}[1]{\ensuremath{y_{#1}}}
\newcommand{\yp}[1]{\ensuremath{y_{#1}^{+}}}

\newcommand{\ngc}[1]{NGC~#1}
\newcommand{\sgr}[1]{Sgr\thinspace #1}
\newcommand{\gsim}{\ensuremath{\gtrsim}}
\newcommand{\lsim}{\ensuremath{\lesssim}}
\begin{document}

\shorttitle{$^{3}$He in the Milky Way}
\shortauthors{Bania et al.}

\title{$^{\bf 3}$He in the Milky Way Interstellar Medium: Ionization Structure}

\author{T. M. Bania\altaffilmark{1}, Dana S. Balser\altaffilmark{2},
Robert T. Rood\altaffilmark{3}, T. L. Wilson\altaffilmark{4}, \& 
Jennifer M. LaRocque\altaffilmark{5}}

\altaffiltext{1}{Institute for Astrophysical Research, Department of Astronomy,
Boston University, 725 Commonwealth Avenue, Boston MA 02215, USA. 
(bania@bu.edu)}
\altaffiltext{2}{National Radio Astronomy Observatory, 520 Edgemont Rd., 
Charlottesville, VA 22903, USA.}
\altaffiltext{3}{ P.O. Box 400325, Astronomy Department, University of Virginia, 
Charlottesville VA 22904-4325, USA.}
\altaffiltext{4}{ESO Room 422, Karl-Schwarzschild-Str. 2, 85748
Garching, Germany}
\altaffiltext{5}{Saint Michael's College, Colchester VT 05439}

\received{}
\revised{}
\accepted{}
\cpright{AAS}{2007}

\begin{abstract}
The cosmic abundance of the \he3\ isotope has important implications
for many fields of astrophysics.  We are using the 8.665\ghz\
hyperfine transition of \hep3\ to determine the \her3\ abundance in
Milky Way \hii\ regions and planetary nebulae.  This is one in a
series of papers in which we discuss issues involved in deriving
accurate \her3\ abundance ratios from the available measurements.
Here we describe the ionization correction we use to convert the
\hepr3\ abundance, \yp3, to the \her3\ abundance, \y3.  In principle
the nebular ionization structure can significantly influence the \y3\
derived for individual sources.  We find that in general there is
insufficient information available to make a detailed ionization
correction.  Here we make a simple correction and assess its validity.
The correction is based on radio recombination line measurements of
\hp\ and \hep4, together with simple core-halo source models.  We use
these models to establish criteria that allow us to identify sources
that can be accurately corrected for ionization and those that
cannot. We argue that this effect cannot be very large for most of the
sources in our observational sample.  For a wide range of models of
nebular ionization structure we find that the ionization correction
factor varies from 1 to 1.8.  Although larger corrections are
possible, there would have to be a conspiracy between the density and
ionization structure for us to underestimate the ionization correction
by a substantial amount.
\end{abstract}


\section{INTRODUCTION}

\subsection{The $^{\bf 3}$He Experiment}\label{sec:he3}

The abundance of the light isotope of helium, \he3, is astrophysically
important.  Knowing the \her3\ abundance ratio can be used to test the
theory of stellar nucleosynthesis; it gives important information
needed to evaluate models of Galactic chemical evolution; it can help
constrain Big Bang Nucleosynthesis.  For over two decades we have used
the 8.665\ghz\ hyperfine transition of \hep3\ to derive the \her3\
abundance in the interstellar medium (ISM) of the Milky Way.  Our
\he3\ sources, planetary nebulae and \hii\ regions, are distributed
throughout the Milky Way's disk, from the Galactic Center to the
outermost regions.  There is no other \he3\ spectral transition
available that can be used to probe transgalactic paths with the
sensitivity and accuracy of the \hep3\ hyperfine line.

We observed \hep3\ using the Max-Planck-Institut f\"{u}r Radioastronomie
(MPIfR) 100\m\ telescope and the National Radio Astronomy Observatory
(NRAO)\footnotemark[1] 140\,Foot telescope and Very Large Array (VLA).
\footnotetext[1]{\nraoblurb} With the 100\m\ and VLA we primarily studied
planetary nebulae \citep{BBRW97, balser06} whereas the 140\,Foot was used mostly
for \hii\ regions \citep[hereafter Paper I]{BBRWW97}.
\defcitealias{BBRWW97}{Paper I}
In addition to the \hep3\ observations, measurements were also made of
a variety of radio recombination line transitions (RRLs) of H, \he4,
and C, as well as of the thermal continuum emission from the nebulae.
Paper I discussed the observations and the measurement errors.  It is
a summary of the status of the observations as of 1996 March.  We
continued the \hep3\ program on the 140\,Foot until it was
decommissioned in 1999 July.  (The last scientific spectrum observed
at the 140\,Foot telescope was the \hep3\ scan that finished at
08:12:20 EDT 19 July 1999.) The final paper in this series
\citep[Paper IV]{RBB04} summarizes the results of the \hep3\ 
experiment for all the sources observed with the NRAO 140\,Foot
telescope during the period 1982--1999.  Paper IV compiles the
observed properties of the \hep3\ emission and gives the final \her3\
abundances derived for the NRAO 140\,Foot sample of
\hep3\ \hii\ regions.

The quantity of astrophysical interest is the \her3\ abundance ratio by
number which we define as \y3.  This ratio provides information about
both stellar \citep{Charbonnel98} and Galactic chemical evolution
\citep{Romano03} and constrains cosmological models during the era of
primordial nucleosynthesis \citep{Yang84, BRB02}.  The species directly
accessible to observation, however, are \hep3\ and \hp.  The
collisionally excited \hep3\ hyperfine transition directly measures the
total column density of \hep3\ atoms along the line-of-sight within the
telescope's beam.  To determine the column density of \hp, either radio
continuum or hydrogen RRLs can be used.  At 8.7\ghz\ the radio continuum
emission in \hii\ regions and planetary nebulae is primarily due to
thermal free-free emission which is proportional to the emission
measure, $EM = \int \, n_{e}^{2} \, d\ell$, where $n_{e}$
is the electron density and $d\ell$ is the differential path length
through the ionized nebula.  The H RRL emission is also proportional to
the emission measure.  Thus neither the RRL nor the radio continuum data
for H$^{+}$ are probing the total proton column density (or mass) in a
straightforward manner.  {\it In order to determine the \hepr3\
abundance ratio, it is therefore necessary to model the nebula's density
structure.\/} Models are required because in most cases detailed
information on nebular density structure is not observationally
accessible.  Nebular density models and the \hepr3\ abundance ratios
derived from them are discussed by \citet[hereafter Paper II]{BBRW99}.
\defcitealias{BBRW99}{Paper II}
In particular, Table 5 of Paper II lists the adopted \hepr3\ abundance
ratios for the 21 \hii\ regions discussed in Paper I.

Using the observables, \hep3\ and \hp, to derive a \her3\ abundance
ratio, \y3, is, however, a two step process. The first step is to
determine the source density structure and calculate \hepr3\ (Paper
II).  The next step, the topic of this paper, is to determine the
source ionization structure and to use this information to convert
\hepr3\ into \her3.  This requires an understanding of the ionization
properties of each ionized nebula.  Specifically, one needs to know
the ionization structure of both the He and H throughout each nebula.
Because the first ionization potentials of H (13.6 eV) and He (24.6
eV) are nearly a factor of two different, in principle ionization can
significantly influence \y3\ for individual sources.  We find that in
general there is insufficient information available to make a detailed
ionization correction.  Here we make a simple correction and assess
its validity.

\subsection{Helium Ionization Structure in H\,{\bf\small II} Regions}
\label{sec:struc}

Several diagnostics are commonly used to study the helium ionization
structure in \hii\ regions.  The most direct is to measure the
fraction of \heo4, \hep4, and \hepp4\ within the \hii\ (${\rm H}^{+}$)
region.  Since both \he4\ and \he3\ have essentially the same
ionization potential the \hep3\ and \hep4\ emission should come from
identical zones within the nebula.  The \hep4\ and \hepp4\ emission
can be measured using recombination lines.  Unfortunately there is no
direct way of measuring the amount of neutral helium within the \hii\
region.  In some cases the spectral types of the ionizing OB stars can
be identified from optical photometry and spectroscopy.  The helium
ionization properties are then derived using stellar atmosphere models
to calculate the expected escaping flux \citep[e.g.,][]{Vacca96},
together with photoionization models of the nebula to determine the He
ionization structure \citep[e.g.,][]{Rubin84}.  At optical wavelengths
spectral transitions of several other atomic species are also observed
to help constrain these models.  The most detailed analyzes have been
made for the Orion nebula in order to determine the total \her4\
abundance ratio \citep{Mathis91, Baldwin91, Rubin91, Pogge92,
Esteban98, Blagrave06}.  Nevertheless, the limiting factor in
determining accurate total helium abundances in Orion is in converting
\hepr4\ to \her4.

Most \hii\ regions are ionized by several stars and may thus have a
complex geometry. Because of this little direct information may be
known about the radiation field.  A variety of diagnostics have been
developed to probe the ionization structure.  \citet{Vilchez88}
developed a radiation softness parameter based on the fine structure
lines of O and S ions that is not very sensitive to chemical
composition \citep[see also][]{Shields78, Mathis82,
Mathis85}. \citet{Armour99} use infrared transitions of Ne and Ar ions
to determine the ionization structure.  In general \hii\ regions
ionized by a hard radiation field produce the most accurate \her4\
abundance ratios since all of the helium within the \hii\ regions will
be ionized.  For example, metal poor blue compact galaxies with \hii\
regions ionized by hard radiation fields are used to measure
primordial \her4.  The ionization structure of these objects has been
extensively studied \citep{Ballantyne00, Viegas00, Gruenwald02,
Sauer02, Peimbert02a, Izotov07}.  In some cases the total \her4\
abundance ratio will be less than \hepr4\ by a few percent,
significant for cosmological implications, since the \heii\ zone is
larger than the \hii\ zone.  If clumping exits in these \hii\ regions,
however, the determined \y4\ values will be underestimated by a few
percent \citep{Mathis05}.  In the Galaxy the high excitation \hii\
regions M17 and S206 have been used instead of Orion to determine the
total \her4\ abundance ratio 
\citep{Peimbert93, Esteban99, Deharveng00, Balser06}.

Many of our sources are located throughout the Galactic disk and are
totally obscured at optical wavelengths by dust.  Furthermore, most of
the optically visible \hii\ regions in our sample are low emission
measure \hii\ regions in the outer Galaxy.  The low EM renders useless
the Orion-type detailed models since the spectral diagnostics used to
constrain them are not available.  Fortunately, unlike many of the
light elements, \her3\ abundances accurate to $\sim$\,10\% can yield
important astrophysical conclusions \citep{Wilson94} .  Because of
this we adopt here a simple ionization structure model that is
constrained primarily by H and \he4\ radio recombination line
observations.  We then use numerical models to assess the accuracy of
this approach for the \hii\ regions in our sample.  Our goal is to
identify a subset of sources in our nebular sample wherein we can
derive \her3\ abundances accurate to $\sim$\,10\%.

\subsection{${\bf {}^{\bf 3}He}$ Ionization Correction}
\label{sec:defkap}

We seek an ionization correction factor to derive the \her3\ abundance
from the \hepr3\ abundance gotten from the \hep3\ observations.  We
define the ionization correction factor, \kapi, to be
$y_{3} \equiv \kappa_{i}\,y_{3}^{+}$, where $y$ is the He/H
abundance ratio by number, the subscript denotes the isotope, and the
superscript is the ionization state.\footnotemark[2]
\footnotetext[2]{Thus the notation used throughout is: \y3\ = \her3;
\yp3\ = \hepr3; \y4\ = \her4; and \yp4\ = \hepr4.}
The ionization correction factor is determined by using the H and
\he4\ radio recombination line observations.  We assume that the \hep3\
emission traces the \hep4\ emission, that the amount of neutral and
doubly ionized helium is negligible, and a canonical value for the
total \her4\ abundance, \y4. The \her3\ abundance ratio is then given by
\begin{equation}\label{eq:defkap}
{ 
y_{3} = \biggl(\frac{y_{4}}{y_{4}^{+}}\biggr)\,y_{3}^{+} 
= \kappa_{i}\,y_{3}^{+}\,\biggl(\frac{y_{\rm 4GAL}}{0.10}}\biggr).
\end{equation}
Here $y_{\rm 4GAL}$ is the actual \he4/H abundance by number in the
Milky Way.  There are, of course, some problems with this simple
model.  \hep3\ and \hep4\ are measured using transitions that are
sensitive in different ways to density structure within the nebula.
The \hep3\ is observed using a collisionally excited hyperfine
transition that is sensitive to the column density and thus
proportional to the electron density, $n_{e}$.  The \hep4\ is observed
using recombination transitions that are proportional to the emission
measure or $n_{e}^{2}$.  Thus if the \hii\ region has large density
fluctuations the different transitions may probe significantly
different material.  For example, consider a simple two-component
nebula with a small, very dense core and a larger, diffuse halo.
Under certain physical conditions the halo will dominate the \hep3\
emission while the core will dominate the \hep4\ emission.  If this is
coupled with ionization structure between the core and halo then the
simple formula in equation (1) will be incorrect.

It has also proven difficult to measure accurately the \her4\
abundance, \y4, in the Galaxy. \he4\ cannot be directly measured in
the Sun and must be inferred from theoretical stellar evolution models
and helioseismology.  Measurements of \he4\ in \hii\ regions using
recombination lines require an ionization correction as discussed
above.  A canonical value of $\y4 = y_{\rm 4GAL} = 0.1$ is adopted
here although there is evidence that this often cited value may be too
high (see \S\ref{sec:discuss}).  For this reason we have parameterized
equation (1) with the $y_{\rm 4GAL}$ factor so that our ionization
corrections can be easily scaled should an accurate Milky Way \y4\
value be derived in the future.

\subsection{Radio Recombination Line Observations}\label{sec:rrl}

Here we use high signal-to-noise ratio recombination line spectra
taken at two different spatial resolutions to probe for any
significant ionization structure.  For our \hii\ region sample we have
obtained H and \he4\ RRL data simultaneously for several different
transitions near 8\ghz\ with a spatial resolution of 3\farcm5 (see
Paper I).  These observations are briefly reviewed here in
\S\,\ref{sec:Xband}.  Because they sample the same \hii\ region 
zone as the \hep3\ spectra, these data are primarily used to determine
\yp4\ in equation (1).  Most \he4\ RRL data are from single-dish
telescopes which typically have spatial resolutions \gsim\ 1\arcmin.
Measuring accurate \yp4\ abundance ratios with single-dish telescopes
is difficult because non-random frequency structure is produced in the
instrumental baselines owing to reflections from various parts of the
telescope structure \citep[see][]{Lockman82}.  Improvements in
receiver technology have provided better stability and enhanced
signal-to-noise.  This, together with a better understanding of the
instrumental effects, has enabled more accurate measurements of \yp4\
(Peimbert et al. 1992a; Paper I).

Using these improved techniques we made additional observations at
18\ghz\ with a spatial resolution of 1\farcm5 for a subset of our \hii\
region sample (\S\,\ref{Kband}).  The smaller beam of the 18\ghz\ RRLs
typically probes the more compact, dense components, whereas the larger
3\farcm5 beam of the 8\ghz\ RRLs is more sensitive to the extended,
diffuse material.  Radio interferometers can measure the \hep4\
distribution within \hii\ regions at spatial resolutions \lsim\
1\arcmin.  Although interferometers do not detect the diffuse emission
because of the zero-spacing flux problem, this information can be
provided by single-dish measurements.  To our knowledge only five \hii\
regions in our sample have been observed in H and \he4\ RRL emission
with radio interferometers: W3, \sgr{B2}, W43, W49, and W51.

Since the entire nebula can be imaged, the best optical data for this
study are observations made with Fabry-Perot spectrophotometers.
\citet{Caplan00} derived oxygen and helium abundances for 34 \hii\
regions using the ESOP Fabry-Perot instrument.  For many objects the
entire nebula was probed (the largest diaphragm was 4\arcmin\
27\arcsec).  The \he4\ abundances from this survey are discussed by
\citet{Deharveng00}.  There are four \hii\ regions that are in our sample:
S206, S209, S212, and S252.

Although the 8\ghz\ RRLs are primarily used to determine \yp4\ the other
radio and optical \hep4\ data can not only provide an independent
confirmation of the \hepr4\ abundance, but also indicate whether the
assumptions in equation (1) are valid.  In \S\,\ref{sec:models}
numerical models are used to explore the physical conditions under which
the assumptions in equation (1) begin to fail.  This information is then
used in \S\,\ref{sec:kappa} to determine a value for the ionization
correction factor, \kapi, for each source. A discussion of the results
is in \S\,\ref{sec:discuss} and a summary of the paper in
\S\,\ref{sec:summary}.  

\section{$^{\bf 4}$He$^{\bf +}$ OBSERVATIONS}\label{sec:data}

\subsection{X-band (8\,GHz) Radio Recombination Lines}
\label{sec:Xband}

The X-band RRL observations are discussed in \S{2} of Paper I.  The
\hal\ and \heal\ transitions were observed in every \hii\ region.
Additional, higher order, transitions were also observed at nearby
frequencies with essentially the same spatial resolution (3\farcm5).
These data were used to constrain the nebular models in Paper II.
Here we only use the higher quality 91$\alpha$ and 114$\beta$
transitions of H and \he4.

Calibration details are discussed in \S{2.1} of Paper I.  We used the
planetary nebula \ngc{7027} to calibrate the flux density scale for
each observing epoch.  The absolute calibration is judged to be within
$\sim$\,10\%; the internal consistency is $\sim$\,5\%.  From these
observations we use only the ratios of the line areas to calculate
\hepr4; the calibration is therefore not a significant source of
uncertainty.

A much larger source of uncertainty are the standing waves produced by
reflections from various parts of the telescope structure (see \S{2.2}
in Paper I and references therein).  To minimize the amplitude of
these standing waves, observations are made with focus offsets
alternating between $\pm \lambda_0/8$ every 60\s.  Cancellation of the
standing waves is, however, not perfect and the spectral baselines
must be modeled.  A 12$^{\rm th}$ order polynomial baseline is 
used to model this non-random frequency structure.  The necessity for
using a high-order baseline and the selection of the appropriate order
to be used is discussed in \citet{Balser94} and Paper I.  The
results of Gaussian fits to each transition are given in Table 4 of
Paper I.

\subsection{K-band (18~GHz) Radio Recombination Lines}
\label{Kband}

Observations of H and \he4\ RRLs at K-band were made with the NRAO 140
Foot telescope during the periods: 1995 November, 1995 December, and
1996 January.  The 140 Foot telescope has a half-power beam-width of
1\farcm5 at 18\ghz.  The radiometer consisted of a dual circularly
polarized HEMT receiver located at the Cassegrain focus of the 140 Foot
telescope.  The system temperature on cold sky under good weather
conditions was $\sim 50$\kel.  The spectrometer consisted of the NRAO
model IV 1024 channel autocorrelator (AC).  The AC sampled four
independent 20\mhz\ wide frequency bands (``quadrants'') of 256 channels
each.  The velocity resolution at 18\ghz\ is 1.3\kms.  The 20\mhz\
bandwidth allowed both the H and \he4\ transitions to be observed
simultaneously within each quadrant.  For most of the observing the AC
was configured with the \halk/\healk\ and the \hbetk/\hebetk\ RRLs in
two quadrants each, simultaneously sampling orthogonal polarizations in
order to maximize our spectral sensitivity.  In some of the brighter
objects the 100$\gamma$ and 110$\delta$ transitions of H and \he4\ 
were observed.

Spectra were taken using the total power observing mode.  First a
reference spectrum (OFF) was taken at a position offset 6 minutes in
right ascension from the nebula and then the nebula itself (ON) was
observed.  Each position had an integration time of 6 minutes.  Local
pointing corrections were determined every hour.  After each pointing
a continuum cross scan, in right ascension and declination, was made
on the \hii\ region to measure the radio continuum emission.

The radiometer system temperature was measured using noise tubes
calibrated by connecting the receiver to matched hot and cold loads.
An ambient temperature absorber was used for the hot load and the
zenith sky for the cold load.  Since this procedure has many
uncertainties and is not always performed before each observing
session we have re-calibrated the system temperature by using the
observed continuum intensity of the unresolved planetary nebula
\ngc{7027}.  \ngc{7027} was observed several times during each
observing session during good weather near transit; it has a flux
density of 6\jy\ at 18\ghz\ \citep{Peng00}.

There are two effects that can modify the calibration as a function of
time.  First, at these frequencies the aperture efficiency of the 140
Foot telescope changes with elevation as gravity distorts the primary
reflector; an elevation gain correction must therefore be made.  We
measured the \hii\ region continuum intensity every hour for each
source in order to determine the appropriate gain correction.  Second,
the atmospheric opacity is significant and depends on the water vapor
content.  We made a tipping scan several times during each observing
period to measure the opacity.  Since we are interested in calculating
the \hepr4\ abundance ratio, which depends on the ratio of the line
intensities, the calibration effectively cancels.  All intensities
quoted here are therefore antenna temperatures calibrated using only
the \ngc{7027} data.

We used the techniques discussed in \S{2.2} of Paper I to model the
K-band spectral baselines, except that a 6$^{\rm th}$ order polynomial
was used instead of a 12$^{\rm th}$ order.  Although the AC
configuration was the same, yielding a 20\mhz\ bandwidth with 256
channels, only $\sim 12$\mhz\ is required to cover the frequency range
between the H and \he4\ lines.  We have therefore reduced the total
number of terms in the polynomial baseline model.

\subsection{The ${\bf {}^{\bf 4}He^{\bf +}\,/\,H^{\bf +}}$ Abundance Ratio}
\label{Yplus}

\begin{figure}
\includegraphics[angle=-90,scale=0.65]{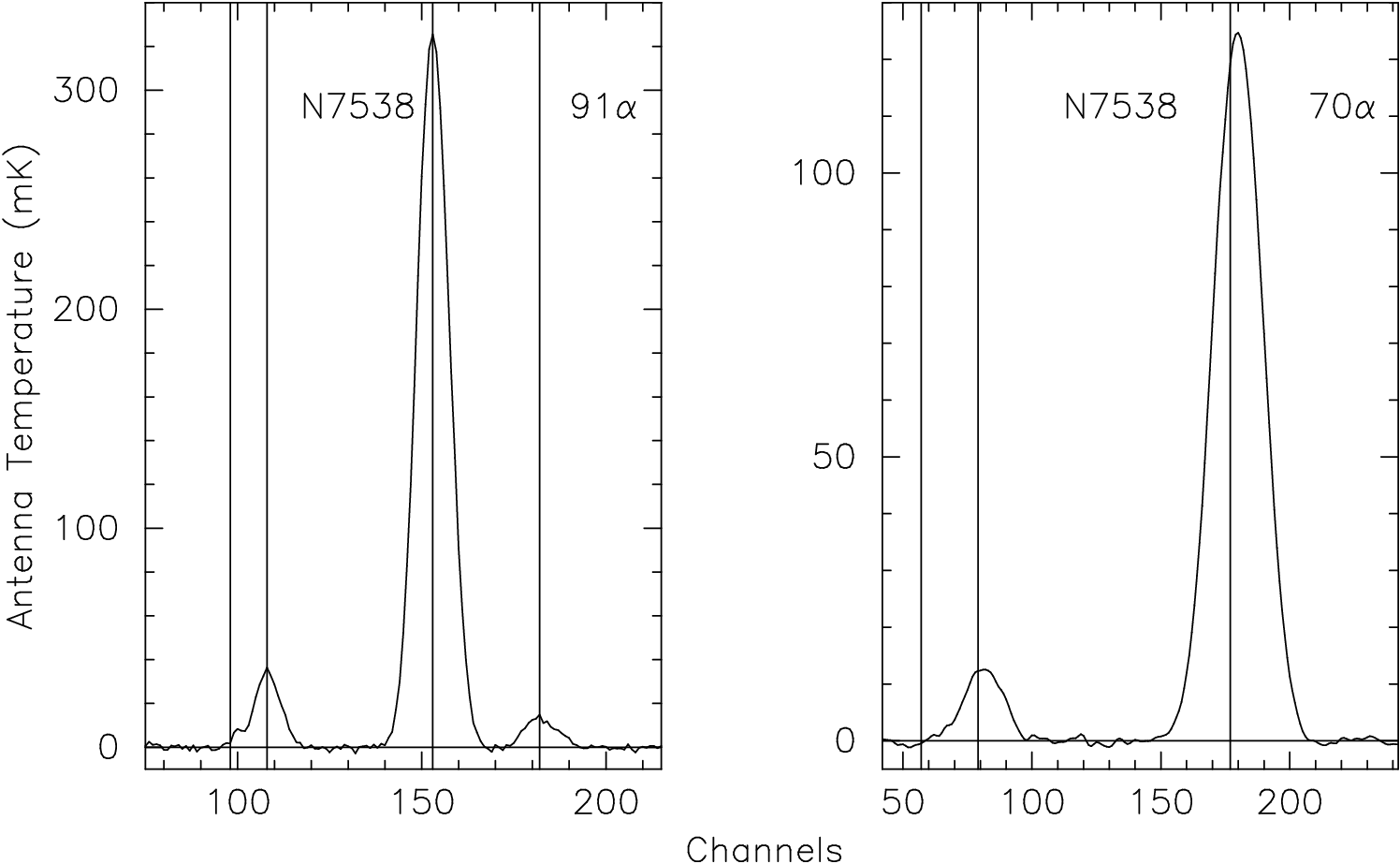}
\includegraphics[angle=-90,scale=0.65]{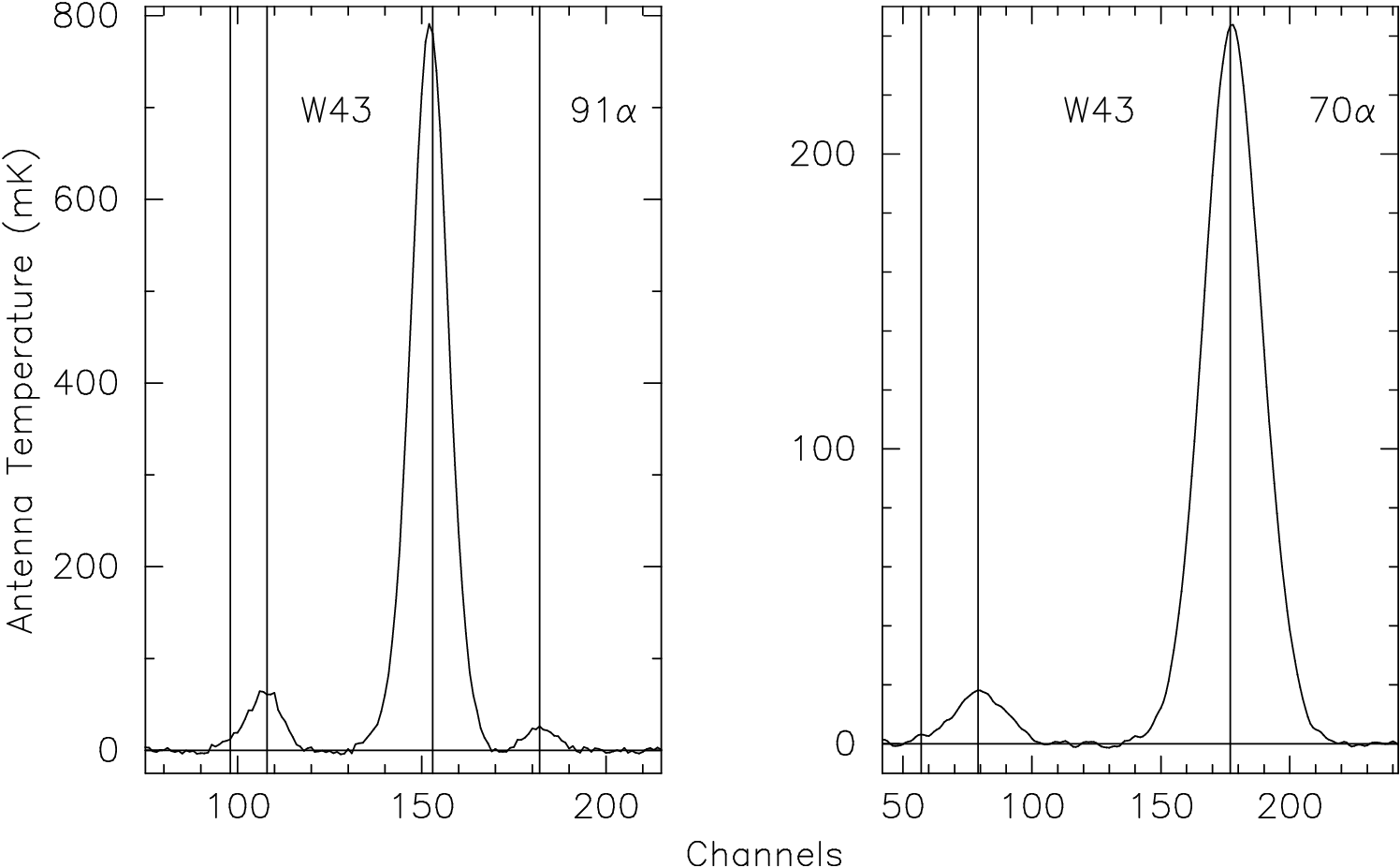}
\caption{Spectra of the 91$\alpha$ X-band (8\ghz) and the
70$\alpha$ K-band (18\ghz) radio recombination line emission for the
\hii\ regions \ngc{7538} and W43.  The vertical lines flag from
left to right the C, He, and H transitions.  (The additional flag in
the 91$\alpha$ spectra is the H\,154$\epsilon$ transition.)  The
intensity scale is in units of milliKelvins.  The LSR velocity
increases from left to right at 1.3\kms\ per channel.
\label{fig:h90h70}}
\end{figure}

Example RRL spectra are shown in Figure~\ref{fig:h90h70}.  Only 10
objects from the original sample of 21 were observed at K-band.  The
data were smoothed to a velocity resolution of 8.1\kms.  The vertical
lines flag the expected locations of the H, He, and C lines for the
specified RRL transition (91$\alpha$ or 70$\alpha$).
Table~\ref{tab:data} summarizes the results of Gaussian fits to the
K-band spectra. Listed are the source name, the RRL transition, the peak
intensity and full-width half-maximum (FWHM) line-width together with
their associated errors, the total integration time, the resulting
r.m.s.  noise of the spectral baseline, and the quality factor of the
line.  The quality factor is determined using the same criteria as in
Paper I.  These include the signal-to-noise ratio, the structure of the
baseline, and the crowding of spectral lines.  We do not include any
calibration error in assessing the quality factor since we only use line
ratios in this paper.

\begin{figure}
\includegraphics[angle=-90,scale=0.60]{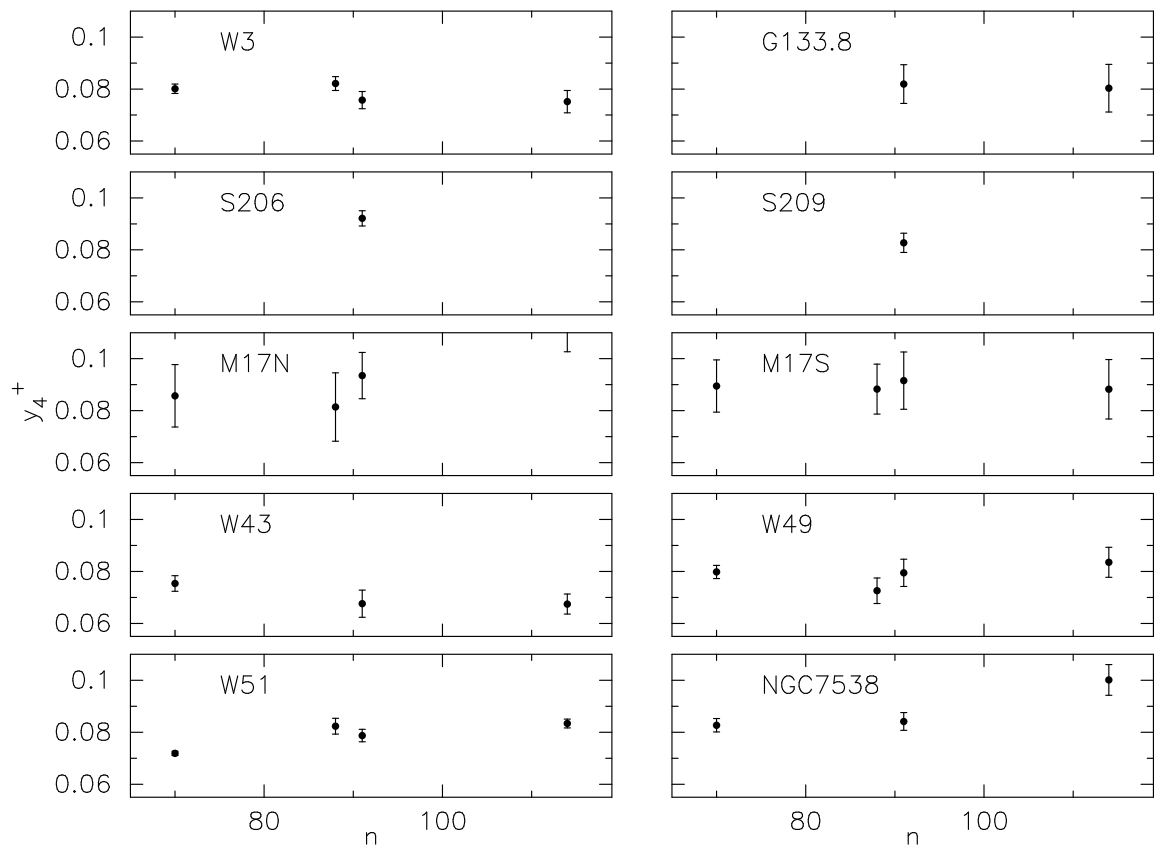}
\caption{The \hepr4, \yp4, abundance ratio plotted as a function of the
principal quantum number, $n$, of the RRL transition.  The 70$\alpha$, 
88$\beta$, 91$\alpha$, and 114$\beta$ transitions are shown.  
Abundances are plotted only for transitions that have a quality
factor of A or B (see Table I and Paper I).
\label{fig:QMn}}
\end{figure}

The \hepr4\,, \yp4, abundance ratios calculated from the RRL data using
the line areas are summarized in Figure~\ref{fig:QMn} which plots the
\yp4\ abundance ratio as a function of the principal quantum number,
$n$, of the RRL transition.  The 70$\alpha$, 88$\beta$, 91$\alpha$, and
114$\beta$ transitions are shown.  The Gaussian line parameters in Table
4 of Paper I and in Table 1 here are used for the X-band and K-band
transitions, respectively.  The errors are determined by propagating the
uncertainties in the Gaussian fit to the line intensity and width.

\begin{figure}
\includegraphics[angle=-90,scale=0.60]{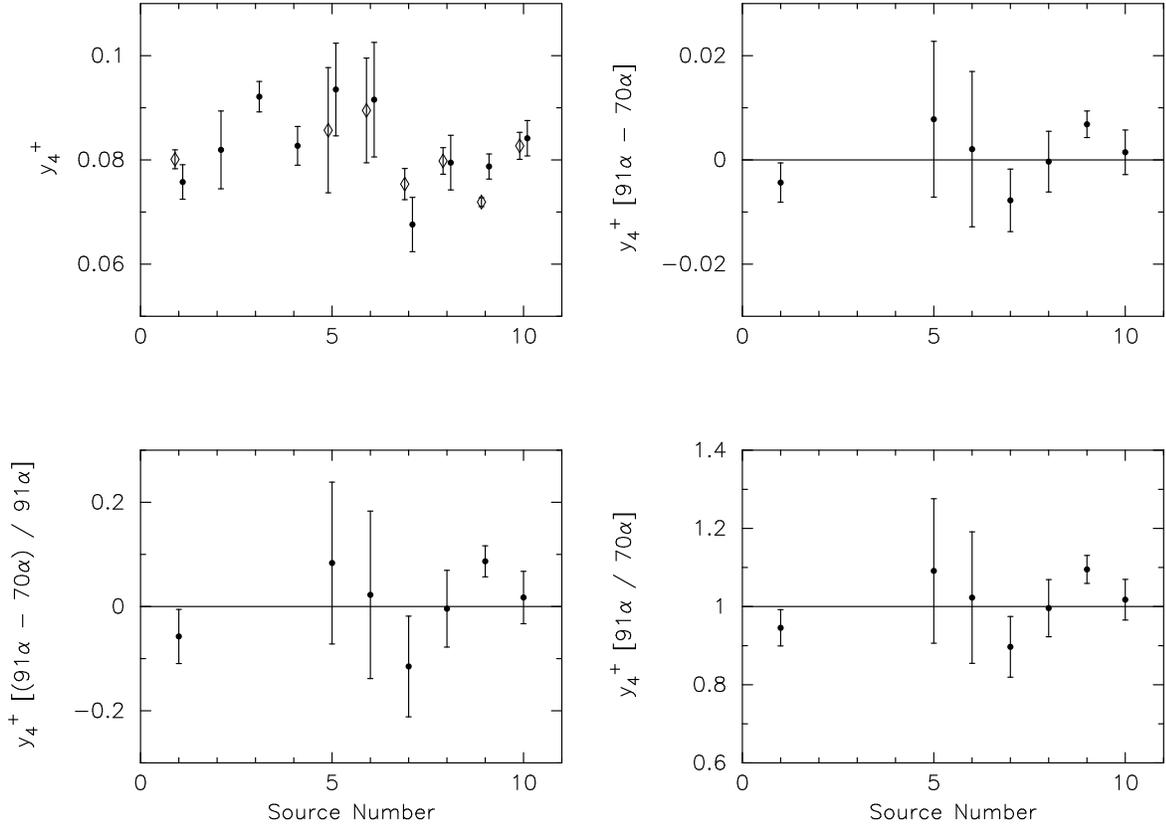}
\caption{Comparison of the \yp4\ ratios derived from the \hal\ 
and \halk\ spectra for the 10 K-band sources in the order listed in
Table~\ref{tab:data}.  Only abundances for quality factor A or B are
plotted.  {\it Top Left:\/} The
\hal\,(circles) and \halk\,(diamonds) \yp4\ ratios for each nebula.
{\it Top Right:\/} The \yp4\,(\hal)\,$-$\,\yp4\,(\halk) difference.
{\it Bottom Left:\/} The fractional \yp4\ deviation:
(\yp4\,(\hal)\,$-$\,\yp4\,(\halk))\,/\,\yp4\,(\hal).  {\it Bottom
Right:\/} The \yp4\,(\hal)\,/\,\yp4\,(\halk) ratio.
\label{fig:yplus}}
\end{figure}

For some nebulae there are real differences among the \yp4\ ratios
derived from the various RRL transitions.  This is to be expected, of
course, since the K-band spectra probe, on average, the denser, more
compact nebular interior whereas the X-band spectra measure a much
larger region, perhaps including a low density halo.  These
statistically significant differences are clearly seen in
Figure~\ref{fig:yplus} which compares in various ways the \yp4\ ratios
derived from the \hal\,(diamonds) and \halk\,(circles) spectra for the
10 K-band sources (see \S\,\ref{sec:discuss}).

\section{SYNTHETIC NEBULAR MODELS}\label{sec:models}

We use the formalism defined in \S\ref{sec:defkap} together with
numerical models for the nebulae to explore the range of validity of
this approach to the ionization correction.  The \hii\ regions in our
sample span a wide range of physical conditions. At one extreme there
are large, complex \hii\ regions such as W49 that are ionized by many
OB-type stars and have multiple compact components in their density
structure.  At the other extreme are \hii\ regions such as S206, which
are smaller, much less complex sources, that are probably ionized by a
single OB-type star.  The simple ionization structure correction model
discussed in \S\,\ref{sec:defkap} could well not be very accurate for
sources like W49.  It is important, however, to understand under what
conditions this model breaks down and what impact this has on deriving
\her3\ abundance ratios.  

Paper II explored the effect of source density structure on the
derivation of the \her3\ abundance.  We used RRL and continuum data to
constrain the models.  Both RRL and continuum emission intensities are
proportional to $n_{e}^2$ and so they trace the densest components
of the nebulae.  We found that our sources could be grouped into two
broad categories.  In one class were ``complex,'' W49-like sources
that required models with complex density structure and non-LTE
radiative transfer in order to reproduce the observations.  In the
other class were ``simple,'' S206-like sources whose observations
could be reproduced by models with LTE radiative transfer and
homogeneous density structure.

We adopt a simple core-halo model for the nebular density and ionization
structure.  From the models we calculate simulated observations and
assess the effect of the model structure on our ability to recover
accurately the \her3\ abundance using our simple ionization structure
correction.  We use a computer program called NEBULA to model the \hii\
region (see Balser 1995 and Paper II for details about NEBULA). The
model nebula can have arbitrary density, temperature, and ionization
structure.  The gas consists of only hydrogen and helium.  We calculate
the radiative transfer through the model nebula for the radio continuum,
recombination line, and \hep3\ hyperfine line emission.  Since we shall
be comparing these synthetic spectral lines to the \S\,\ref{sec:data}
data taken at 8 and 18\ghz, we calculate the model antenna temperature
for each species by convolving the model sky brightness temperature
distribution with the telescope beam and multiplying by the telescope
main beam efficiency.  The telescope beam is assumed to be a Gaussian
with a half-power beam-width (HPBW) of 1\farcm5 or 3\farcm5.  These are
the HPBW sizes of the 140\,Foot telescope beam at 18\ghz\ and 8\ghz,
respectively.

All model nebulae are located at a distance of 8\kpc\ which is roughly
the average distance to our \he3\ sources.  The models are comprised
of nested, spherical, homogeneous, isothermal components.  For
simplicity, all model components have a fixed, 8,000\K\,, electron
temperature that is typical for Galactic \hii\ regions.  For all
models it is assumed that the total abundance ratios are
$\y3 \equiv \nexpo{1}{-5}$ and $\y4 \equiv 0.10$.
Table~\ref{tab:NEBULA} summarizes the model parameters used as input
for the NEBULA program.

Model A is a one-component, low density, $n_e = 100\percc$, nebula
with an angular size of 3\farcm5 (8.14\pc).  All other models are
two-component (core-halo) nebulae which consist of a 1\farcm5
(3.49\pc) core embedded in a larger 3\farcm5 nebula.  Listed for each
component is the angular size, the electron density, the fraction of
the component \hii\ mass relative to the total mass, the \yp3\
abundance ratio, and the \yp4\ abundance ratio.  Altogether, the
models span a core-halo density contrast ratio from 1:1 to 5:1, and a
core mass fraction from 8\% to 30\%.  That is, all models have
halos that contain more mass than the cores by a factor of either
$\sim 2$ or $\sim 10$.  These parameters span the range of the
Galactic \hii\ region physical properties derived from single-dish RRL
observations \citep[e.g., ][]{Quireza06,Quireza07}.

The NEBULA code calculates the following observable quantities: the
continuum antenna temperature at 8.665\ghz\ and the antenna
temperatures of the \hep3, \hal, \heal, \halk, and \healk\
transitions.  Properties of the synthetic observations that result
from the Table~\ref{tab:NEBULA} models are summarized in
Table~\ref{tab:models}.  Listed are the continuum and line antenna
temperature for the \hep3\ transition, the \yp3\ (\hepr3) abundance
ratio, the \yp4\ (\hepr4) abundance ratio and the values of \kapi\ and
\y3\ calculated using the equation (\ref{eq:defkap}) Ansatz.  All of
these quantities are determined for X-band (8\ghz).  The last column
lists the \yp4\ model values at K-band (18\ghz).  The \yp4\ values are
calculated from the H and \he4\ line parameters produced by NEBULA;
they are the ratio of the line areas.
The \yp3\ value is derived from the NEBULA models assuming a single,
homogeneous, fully ionized sphere where ${y}_{3}^{+} \propto
T_{\rm L}\,\Delta{v}/T_{\rm C}^{1/2}$.  This analytical expression 
approximates the numerical calculation to within a few percent 
\citep[see][]{Balser95}.  This is our standard way of
deriving the \her3\ abundance ratio (Paper II).  Here we apply this
analysis to the Table~\ref{tab:NEBULA} nebular models and use the
formalism developed in \S\ref{sec:defkap} to derive the ionization
correction factor, \kapi.

Model A1 is a uniform nebula with no density or ionization structure.
The NEBULA model abundances in Table~\ref{tab:models}, \yp3\ and \yp4,
exactly match the input abundances, \y3\ = \nexpo{1}{-5} and \y4\ =
0.10.  This of course must be the case since Model A1 is constructed
to match our analysis assumptions.  It is a single, homogeneous, fully
ionized sphere and the NEBULA Model A1 gives an ionization correction of
$\kapi = 1$ as it should.

In Model A2 we simulate a one-component nebula that has 30\% of all
the helium neutral.  For this model the NEBULA \he4\ RRL synthetic
spectra together with equation (\ref{eq:defkap}) give an ionization
correction of $\kappa_i = 1.4$ which in turn produces \y3\ =
\nexpo{1.0}{-5}, the model input abundance.  Thus, for a one-component
model nebula the observables are sufficient to derive ionization
corrections that can accurately recover the input structure.

Model B has a core-halo density structure where the core component has
a higher density by a factor of 5, but the halo has a factor of
$\sim 2.3$ times more mass.  Both core and halo are fully ionized:
Model B has no ionization structure so we cannot derive any correction
factor.  The NEBULA Model B \yp3\ abundance thus underestimates the
true \yp3\ abundance by a factor of 1.3.  In Paper II we discuss
density structure corrections for the \hepr3\ abundance ratio
derivation.  There we model nebular density structure and estimate the
density correction factor for each \hii\ region in our sample.
Applying these techniques to Model B recovers, as expected, the input
\yp3\ abundance.

Models C and D are constant density nebulae with core-halo ionization
structure.  The halo mass in both models is $\sim 11.5$ times that of
the core.  Model C has a fully ionized core and a halo that is
under-ionized by 50\%.  Model D reverses this ionization structure.
That these models have ionization structure is revealed by the
different NEBULA values for the X and K-band \yp4\ abundance ratios
(see Table~\ref{tab:models}).  Since most of the mass resides in these
halos the NEBULA \hepr3\ abundance ratio is much less for Model C with
its under-ionized halo (\yp3\ = 0.57) than for model D (\yp3\ = 0.96).
The NEBULA results give ionization corrections of \kapi = 1.82 and
1.05 for models C and D, respectively.  In both cases
Equation~\ref{eq:defkap} yields \y3\ = \nexpo{1.0}{-5}, the model
input value.  For these two models our simple ionization correction
reproduces the correct \her3\ abundance ratio.

Models E and F combine both density and ionization structure in a
core/halo geometry.  Both have high density cores (5:1 density
contrast) and halo masses that are $\sim2.3$ times that of the
cores.  Model E has a fully ionized core and a halo that is
under-ionized by 50\%.  Model F reverses this ionization structure.
Once again the different NEBULA values for the X and K-band \yp4\
abundance ratios reveal the presence of nebular ionization structure.
For model E the NEBULA \yp3\ value is 0.53, roughly a factor of two
too low.  Yet the NEBULA \yp4\ abundance ratio implies only a small
ionization correction factor, \kapi\ = 1.18, which gives a corrected
\y3\ abundance of only \nexpo{6.2}{-6}, 62\% of the input value.
The situation is much better for Model F with its fully ionized halo.
The NEBULA values for \yp4\ and \yp3\ together give an ionization
correction factor of \kapi\ = 1.56 and a \y3\ abundance that is within
2\% of the input value.  

These models show that the simple ionization correction given in
Equation~\ref{eq:defkap} will be reasonably accurate for nebulae with
core/halo geometries.  Our correction begins to break down whenever the
low density component contains most of the mass and is under-ionized.
This is because the \hep3\ and RRL emission arises from gas located in
different zones of the nebula and these regions do not have the same He
ionization structure.  

In sum, the simple ionization correction given by equation 
(\ref{eq:defkap}) provides a reasonable value for the total \y3\
abundance ratio except for the Model E \hii\ region that has an
under-ionized halo containing a significant fraction of the total
mass.  Clearly more complicated density and ionization structures
exist in real \hii\ regions.  The simple models discussed here
nevertheless illustrate the main issues.  The simple ionization
correction breaks down when ionization structure exists between
significantly different density components.  True variations of \y4\
and \y3\ structure within the \hii\ region will complicate the
analysis.  Nonetheless, because of our large X-band beam that in many
cases covers the entire nebula, emission from dense, compact
components tends to be diluted.  We thus derive an average value of
\y3\ for each object.

\section{IONIZATION CORRECTION FACTOR}\label{sec:kappa}

Here we assess the ionization structure correction for our \he3\
sample of 21 Galactic \hii\ regions.  In Paper II we classify nebulae
into two categories: (i)``simple'' \hii\ regions dominated by
low-density, extended structure that is well fit by a single-component
uniform density sphere model; and (ii) ``complex'' \hii\
regions that have multiple high density components and thus require a
more sophisticated model.  The simple \hii\ regions are: G133.8, S206,
S209, S212, S228, S235, S252, S311, S76, S90, S156, \ngc{7538}, and
S162; the complex \hii\ regions are: W3, \sgr{B2}, G1.1, M17S, M17N,
W43, W49, and W51.  Simple \hii\ regions characteristically have low
emission measure.  For complex \hii\ regions that have significant
ionization structure we will underestimate \kapi\ as discussed in
\S\ref{sec:models}.  We therefore can only derive a lower limit
to the \her3\ abundance for these nebulae.  

We used various techniques to assess the excitation state of the
nebulae.  Table~\ref{tab:excitation} summarizes the excitation
properties for our \hii\ region sample.  Listed for the radio data are
the \hepr4\ abundance ratio, \yp4, derived using only the \hal\ and
\heal\ RRLs, the H-ionizing luminosity, $N_{\rm L}$, determined from the radio
continuum data in Paper I, and the spectral type of a single ZAMS star
that can produce the observed luminosity.  Spectral types are derived
from stellar atmosphere models: both the \citet{Vacca96} and \citet[in
parentheses]{Panagia73} models are listed.  Similar results from optical
data are taken from the literature.  The spectral types are based on
photometry and spectroscopy; the earliest type star thought to ionize
each nebula is listed in Table~\ref{tab:excitation}.  Overall, the radio
and optical spectral classifications are in good agreement.

\begin{figure}
\includegraphics[angle=-90,scale=0.70]{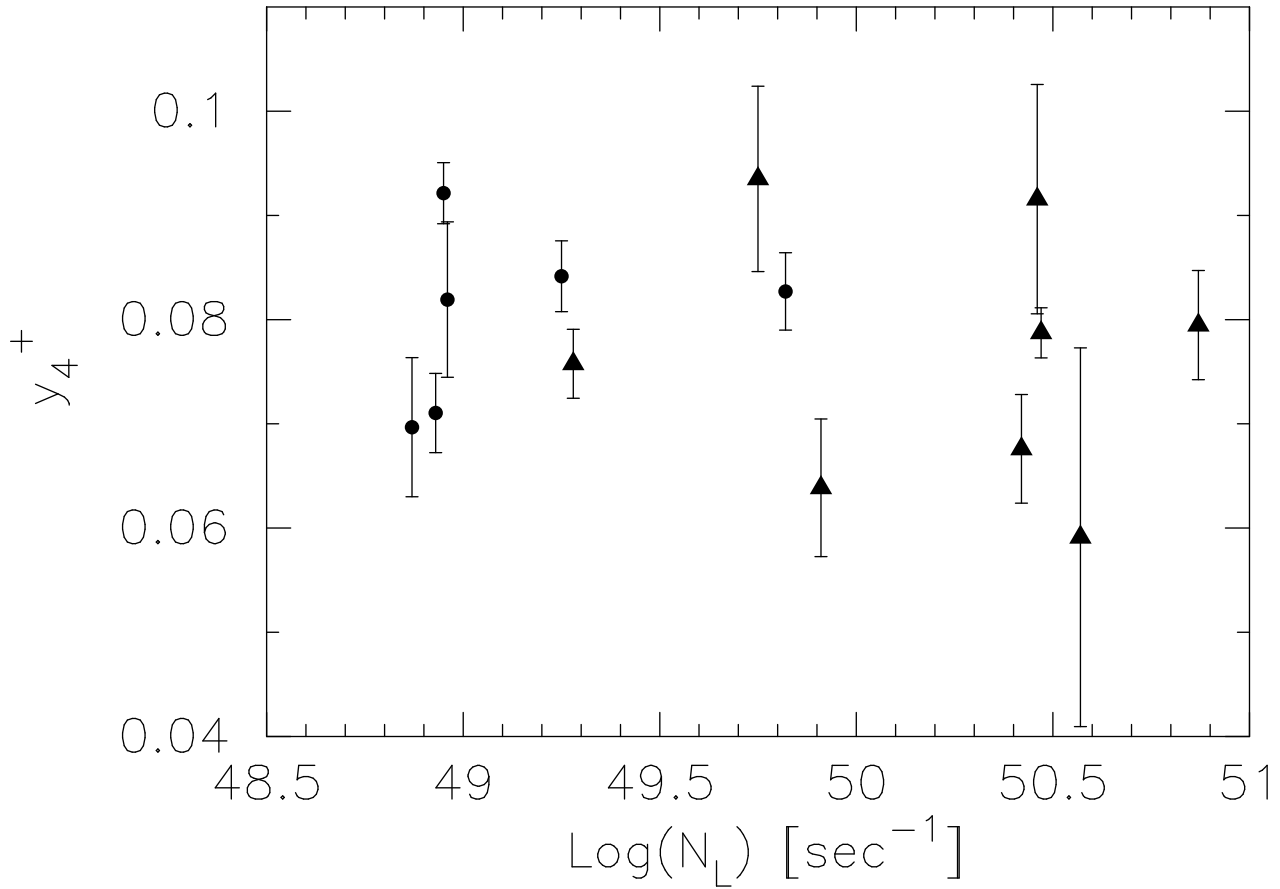} \caption{The
\hepr4, \yp4, abundance ratio plotted as a function of the H-ionizing
luminosity, $N_{\rm L}$.  The \yp4\ values are calculated using only the \hal\
and \heal\ transitions.  Only transitions that have a quality factor of
A or B (see Paper I) are shown.  Circles denote simple sources;
triangles identify complex sources (see text).
\label{fig:Nlc}}
\end{figure}

We plot the ionization state, \yp4, of the 14 nebulae for which we
have sufficient information as a function of the log of the H-ionizing
luminosity, $N_{\rm L}$, in Figure~\ref{fig:Nlc}.  There is no discernible
trend connecting \yp4\ and $N_{\rm L}$.  Except for their systematically
lower $N_{\rm L}$ values (recall they are low emission measure nebulae), the
simple sources span the same \yp4\ range as do the complex nebulae.
Values of $N_{\rm L}$ determined from the radio continuum include the
contribution of H-ionizing photons from all of the early-type stars
within the \hii\ region complex.  Therefore low emission measure does
not necessarily imply that the radiation field is soft. For example
the radiation field of the low emission measure Rosette nebula is
dominated by O4 stars compared to the rather weak ionization field of
the high emission measure Orion nebula.

Below we discuss each \hii\ region separately in more detail.  An
ionization correction factor, \kapi, is determined for each object.
For consistency we only use data as of 1996 March (cf., Papers I and
II).  The \kapi\ values can be multiplied by the \hepr3\ abundance
ratios in Table 5 of Paper II to yield the \her3\ abundance.
\citet{RBB04} derive \her3\ abundances for all the \he3\ nebulae using
the entire 1982--1999 140 Foot dataset.  

{\it W3 ---} Although W3 is optically obscured it has been extensively
studied at radio wavelengths.  Our higher spatial resolution K-band data
give \yp4\ = 0.081 whereas the X-band data yield \yp4\ = 0.076.
Both data sets are of very high quality.  One interpretation of these
results is that the more compact components of W3---in particular
W3A---have a higher excitation than the outer regions probed by the
larger X-band beam.  This is sometimes called the geometric effect where
\yp4\ ratios increase with higher resolution observations.

Interferometers can probe the RRL emission of the nebular gas at high
spatial resolution.  Extant observations include: 76$\alpha$ and
110$\alpha$ data from the Very Large Array (VLA) and the Westerbork
Synthesis Radio Telescope (WSRT) \citep{Roelfsema91} and 92$\alpha$ from
the VLA \citep{Roelfsema92, Adler96}.  The \hepr4\ abundance ratio
varies considerably among the different components ($0.06 \le $ \yp4\ $
\le 0.20$).  Values of \yp4\ higher than 0.10 are thought to be due to
local enrichment from a nearby evolved object.  Alternatively, radiative
transfer effects may cause one to overestimate the true \y4\ abundance
\citep{Gulyaev97}.  High \yp4\ values in other \hii\ regions are
associated with evolved early-type objects \citep[e.g.,][]{Balser01}.
W3 is rather unusual in our sample since the total mass of ionized
gas is small (20$\msun$; paper I) and therefore mass loss from local
objects can significantly pollute the surrounding material.  We
nevertheless adopt $\yp4 = 0.076$ as a reasonable value for most of the
mass in W3.  Because W3 has complex density structure and there is
uncertainty about the ionization structure, we adopt an ionization
correction of $\kappa_i \ge 1.32$.

{\it G133.8 ---} Also known as W3N, G133.8 is the northern most region
of the W3/W4 complex and has a slightly more negative velocity (by 7
\kms) than the other components of W3.  Infrared and radio continuum
observations of G133.8 indicate that this region is ionized by a
luminosity equivalent to an O7 star
\citep[Table~\ref{tab:excitation};][]{Schraml69, Thronson84, Carpenter00}.  
The X-band RRLs measure \yp4\ = 0.082, roughly consistent with the
ionization properties predicted from the continuum observations.  We
adopt an ionization correction of $\kappa_i = 1.22$.

{\it S206 ---} This outer Galaxy \hii\ region is thought to be ionized
by a single star with spectral type O4--O6.  Fabry-Perot
spectrophotometer data show that S206 contains essentially no neutral
helium with \yp4\ = 0.10 \citep{Deharveng00}.  Our X-band measurements
give a slightly lower value of \yp4\ = 0.092.  Recent high sensitivity
RRL data with the Green Bank Telescope yield \yp4\ = 0.085 \citep{Balser06}.  
Since we are correcting 140 Foot \he3\ data, however, we adopt
$\kappa_i = 1.09$.

{\it S209 ---} This \hii\ region has the largest Galactocentric
distance, $\rgal = 16.9$\kpc, in our entire sample and thus is an
important object.  Spectrophotometry reveals an O9 and two B1 stars
exciting S209 \citep{Chini84}; this suggests that S209 might contain
some neutral helium.  Optical \he4\ observations show that \yp4\ 
ranges from 0.084 \citep{Vilchez96} to 0.12 \citep{Deharveng00}.  The
\heal\ transition gives \yp4\ = 0.083.  We adopt \yp4\ = 0.083
producing an ionization correction of $\kappa_i = 1.20$.

{\it S212 ---} \hep3\ was not detected toward S212 and the \he4\ data
are of poor quality.  S212 is nonetheless a potentially important
object with $\rgal = 14.2$\kpc.  The nebula is ionized by an
early-type star, O5.5--O8.  Optical observations measure $\yp4 =
0.088$--0.10; most of the helium should be ionized.  We adopt $\yp4 =
0.088$ which gives an ionization correction of $\kappa_i = 1.14$.

{\it S228 ---} \hep3\ was not detected toward S228 and the \heal\ line
is only a probable detection.  As is the case for S212, this outer
Galaxy nebula is potentially a good \he3\ candidate.  Based on both
radio and optical data, however, this region is ionized by a late-type
O-star; the helium is probably not fully ionized within the \hii\
region.  Because \hep3\ and \hep4\ have not been detected we have a
direct measure of neither the \her3\ abundance nor the ionization
correction.

{\it S235, S76, S90 ---} These physically small, shell-like nebulae were
observed because they have a morphology that is similar to that of
W3\,A.  The anomalously high \her4\ abundance ratio in W3A might be
caused by local enrichment from nearby, evolved objects.  This might
also explain the high W3 \her3\ abundance ratio \citep{Olive95}.

\hep3\ was not detected in either S235 or S76.  Although we do not
have good quality \hep4\ data it is clear that the helium in these
nebulae is significantly under-ionized (Paper I).  Both radio and
optical data show that these two \hii\ regions are ionized by late
O-type or early B-type stars.  Given these soft radiation fields even
a high \her3\ abundance ratio would be difficult to detect.

In S90, both \hep3\ and \hep4\ are detected.  Although an O9.5 star is
optically identified near S90, \citet{Crampton78} argue that this star
does not excite the nebula. The radio continuum data suggest an earlier
type object, O8--O6.5, roughly consistent with the observed value of
$\yp4 = 0.070$.  We adopt an ionization correction factor of $\kapi =
1.43$.

{\it S252 ---} Also known as \ngc{2175} or W13, this diffuse \hii\
region extends over 25\,\arcmin\ on the sky and contains several
compact components.  The nebula is ionized mainly by an O6--O6.5 star
centered on the \hii\ region \citep{Haikala95}.  Our observed position
lies on a bright rim located at the western edge of the nebula
\citep{Felli77}.  Because of the low emission measures in S252 there
is only a probable detection of \heal\ (Paper I).  \hep4\ is, however,
detected at optical wavelengths.  \citet{Shaver83} measured $\yp4 =
0.051$ near our position.  Towards the center of the \hii\ region this
value increases with $\yp4 = 0.091$--0.10 \citep{Shaver83,
Deharveng00}.  We adopt an ionization correction factor of $\kappa_i =
1.96$.

{\it S311 ---} S311 is thought to be ionized by an O6--O6.5 star
located north of the radio emission peak \citep{Persi87}.  The
asymmetric density distribution suggests a blister type geometry
\citep{Albert86}. \heal\ emission is detected: $\yp4 = 0.071$.  We
adopt an ionization correction factor of $\kappa_i = 1.41$.

{\it \sgr{B2} ---} \sgr{B2} is the most complex source in our sample.
It is important because it is located near the Galactic Center (GC).
The ionized gas is observed at high spatial resolution with radio
interferometers \citep[e.g.,][]{Martin72, Benson84, Gaume90, Gaume95}.
There are at least 57 separate, compact components associated with
\sgr{B2} \citep{DePree96}.

The \hepr4\ abundance in \sgr{B2} is known to be anomalous
\citep[see][and references within]{Lockman82}.  The radio continuum
emission implies a very high excitation but the \hepr4\ ratio is at
times reported to be below 0.06 (e.g., Table~\ref{tab:excitation}).  In
the literature the \sgr{B2} \her4\ abundance is reported as
underabundant, overabundant, or normal, depending on various effects
such as the geometric effect, selective photo-absorption by dust, or the
external maser effect.  These uncertainties are compounded by complex
spectral baselines produced by reflections from the telescope structure
for the single-dish observations.

Some insight can, however, be gleaned from high resolution
interferometric observations of RRL emission.  \sgr{B2} was observed
with the VLA in 76$\alpha$ emission \citep{Roelfsema87}; 110$\alpha$
emission \citep{Mehringer93}; and 66$\alpha$ emission \citep{DePree95,
DePree96}.  The \hepr4\ ratio varies from 0.05 to 0.17 among the various
compact components; the average value is 0.10 \citep{Roelfsema87,
DePree96}.  The component F bright cluster of sources is the sole
extremely low \hepr4\ ratio, $\yp4 \le 0.05$, component
\citep{DePree96}.

Most of the radio continuum flux density at X-band arises from the
compact components towards \sgr{B2} \citep{Balser95a}.  The \sgr{B2} F
components have the highest peak emission measures. Since the RRL
emission is proportional to emission measure, the \hep4\ RRL emission is
strongly influenced by the F components.  It is therefore not surprising
that, depending on frequency and spatial resolution, different \yp4\
values are measured in \sgr{B2}.
Because toward \sgr{B2} there is ionization structure coupled with very
complex density structure within the 3\farcm5 beam of the 140 Foot
telescope, using equation (1) is inappropriate.  For this source we can
only adopt a limit to the \her3\ abundance ratio.

{\it G1.1 ---} Better known as \sgr{D}, this \hii\ region is located
only 1\arcdeg\ from the GC but is much less complex than \sgr{B2}.
Although we have placed its location at the GC there is evidence
suggesting that it might be located outside the molecular nuclear disk
\citep{Lis91, Mehringer98, Blum99a}.  High resolution radio continuum
images reveal several compact components within a diffuse, extended
halo \citep{Liszt92, Mehringer98}.  A single O7 star can ionize the
core \citep{Liszt92}.  This is consistent with narrow band imaging at
2.17\micron, Br$\gamma$, and 2.06\micron, and \hei\
\citep{Blum99}.  If the diffuse gas is included in the analysis an O4
star is required to account for the excitation
(Table~\ref{tab:excitation}).  The \hepr4\ ratio of 0.064 implies the
presence of some neutral helium in the nebula.  This is also
consistent with the infrared observations of \citet{Blum99}.
Initially G1.1 was classified as complex because of scant data;
additional RRL data supports a simple homogeneous model.  We therefore
adopt the ionization correction factor of $\kappa_ i = 1.56$.

{\it M17 ---} The Omega nebula, also known as M17, S45, \ngc{6618}, and
W28, is a well studied inner Galaxy \hii\ region.  The nebula spans over
10\,\arcmin\ in angular size and consists of two main components: M17
north (M17N) and M17 south (M17S).  The Omega nebula is complex: high
resolution radio continuum images reveal this density structure directly
\citep{Felli84}.  We were unable to model the density structure in Paper
II.  The complex velocity field produces non-Gaussian profiles in both
M17N and M17S (Paper I; also see Joncas \& Roy 1986).

It is nonetheless clear that most of the helium is ionized.  High
extinction makes optical identification of the ionizing stars difficult,
but infrared spectrophotometry shows at least five O3-O6 stars
\citep{Hanson95}.  The large number of early O-type stars and the high
\hepr4\ abundance ratios observed at both radio and optical wavelengths
suggest that $\kappa_i \approx 1$.  Using the \heal\ transitions we
adopt an ionization correction factor of $\kapi = 1.09$ and 1.06 for
M17S and M17N, respectively.  

{\it W43 ---} W43 is a large, optically obscured \hii\ region located
only 4.6\kpc\ from the GC.  Because \hii\ regions are sparse between
the GC and 4.5\kpc\ \citep{Lockman96}, W43 probes the \her3\ abundance
near the outer edge of this \hii\ region desert.  W43 contains low
density, diffuse gas and many compact components
\citep[e.g.,][]{Subrahmanyan96, Balser01}.  Near infrared observations
show a star cluster at the center of the \hii\ region consisting of a
WN7 star and two O-type stars \citep{Blum99}.  The total H-ionizing
luminosity is considerable, requiring at least an O4 star for the
entire region (Table~\ref{tab:excitation}).

We studied the helium ionization properties of both the diffuse and
compact gas in W43 using the 140 Foot telescope and the VLA
\citep{Balser01}.  Using RRL emission we imaged the \hepr4\ abundance
ratio for both the compact and diffuse components.  The average
\hepr4\ ratio of the entire dataset is $\langle\yp4\rangle = 0.077 \pm
0.01$.  We find no significant variations of the \hepr4\ abundance,
even for positions observed at the edge of the nebula with the 140
Foot.  Thus although W43 is a complex nebula, the ionization
properties do not appear to vary within this object so we can derive
an accurate ionization correction factor.  We adopt $\kappa_i = 1.30$
based on the average \yp4\ abundance.

{\it W49 ---} W49 is one of the most luminous \hii\ regions in the
Galaxy, consisting of a ring of ultra compact \hii\ regions (UC \hii\
regions), each containing at least one O-type star
\citep[e.g.,][]{Welch87}.  These UC \hii\ regions make up only
one-third of the total H-ionizing luminosity observed \citep{Conti02}.
The \heal\ RRL gives a \hepr4\ abundance ratio of $0.079 \pm 0.0052$,
consistent with the K-band data.  Higher resolution VLA H and He RRL
images find \yp4\ ranging from $0.05 \pm 0.02$ to $0.18 \pm 0.06$ in
13 objects with an average of $\langle\yp4\rangle = 0.11 \pm 0.01$
\citep{DePree97}.  This significant ionization structure combined with
considerable density structure makes any equation (\ref{eq:defkap})
type ionization correction inappropriate.  We therefore adopt
$\kappa_i \ge\ 1$.

{\it W51 ---} W51 is a bright, complex \hii\ region with an angular
extent of 2\,\arcdeg\ on the sky.  Radio continuum observations reveal
several O4--O6 stars associated with the main components
\citep{Mehringer94}.  Mid-infrared imaging and spectroscopy suggests
stars with spectral types $\sim$\,O9 \citep{Okamoto01}.  The X-band
\hep4\ RRL data give a \yp4\ abundance of $0.079 \pm 0.0024$.  The
\healk\ transition appears to be slightly lower with \yp4\ $\sim
0.07$.  High resolution VLA observations of 92$\alpha$ RRLs yield
\yp4\ values ranging from $0.08 \pm 0.02$ to $0.14 \pm 0.06$ with
an average of $\langle\yp4\rangle = 0.096 \pm 0.015$ \citep{Mehringer94}.
Although most of the helium appears to be ionized, given the
complexity of this object we adopt an ionization correction factor of
$\kappa_i \ge\ 1$.

{\it S156 ---} S156 is a compact \hii\ region excited by an O6.5-type
star.  We have only an upper limit for the \hepr3\ abundance.  Although
we have detected \hep4\ at X-band the accuracy is not very good.
Optical observations yield $\yp4 = 0.081$ \citep{Deharveng00}.  We adopt
an ionization correction factor of $\kappa_i = 1.23$. 

{\it \ngc{7538} ---} Also known as S158, \ngc{7538} is centrally diffuse
with a bright rim to the west \citep{Israel77}.  An O7 star located at
the center of the nebula is responsible for the excitation
\citep{Deharveng79}.  Optical data yield $\yp4 = 0.10$ \citep{Lynds86}, a
value higher than our RRL results.  Both the X-band and K-band data give
\yp4\ $\sim 0.08$, although the \hebet\ transition produces a slightly
higher value of 0.10.  We adopt an ionization correction factor based 
on the X-band data of $\kappa_i = 1.19$.

{\it S162 ---} Sometimes called the Bubble nebula, S162 consists of a
circularly shaped shell with bright emission to the north observed in
H$\alpha$ and the radio continuum \citep[Paper I]{Barlow76, Israel73}.
The exciting star has been classified as an O6.5-type, consistent with
the radio continuum estimates.  We do not detect \hep3\ emission.
Although we do detect the \heal\ RRL, the quality factor is poor.  We
cannot make an ionization correction so therefore adopt $\kappa_i \ge
1$.

\section{DISCUSSION}\label{sec:discuss}

Our nebular ionization correction, \kapi\,, is the simplest possible
approach to the problem.  It makes several key assumptions: (1) the
\he4\ abundance is \her4\ $\equiv 0.10$ for all Galactic \hii\ regions;
(2) all the He is singly ionized---there is neither neutral nor doubly
ionized He within the nebulae; and (3) the \hep3\ and \hep4\ ions
occupy identical volumes within the nebulae.  Measuring the total
helium abundance is difficult since there is no direct way to observe
neutral helium.  We do not expect any significant doubly ionized
helium since the radiation field from Galactic O-type stars is not
hard enough.  For example, M17 is ionized by several stars classified
between O3--O6 \citep{Hanson95} and is one of the Galactic \hii\
regions with the highest degree of ionization.  Yet only a very low
upper limit, $^3{\rm He}^{++}/{\rm H}^+ < 8 \times 10^{-5}$, is measured for 
\hepp4\ in M17 \citep{Peimbert92b}.  In fact, to our knowledge \hepp4\ 
has never been detected in any Galactic \hii\ region.

A canonical value for the Galactic \he4\ abundance of $\y4 = 0.1$ is
typically used in the literature.  Most of the \he4\ was produced
during the era of primordial nucleosynthesis.  Therefore the variation
of \he4\ in the Galaxy, caused by stellar and Galactic evolution, is
expected to be small because $\Delta Y / \Delta Z \sim 1$, where $Y$
and $Z$ are the helium and metal abundances by mass
\citep[e.g.,][]{Chiappini02}.  Direct measurement of the total \her4\
abundance ratio in the Galaxy is difficult. The \he4\ abundance cannot
be determined using either the Solar photosphere or in meteorites.
Measurements of \he4\ in Galactic \hii\ regions have to correct for
ionization structure just as we are trying to do here for \he3.

Solar \he4\ abundances are determined from theoretical stellar
evolution models with \y4\ $\sim 0.1$ \citep[e.g.,][]{Anders89}.
\citet{Peimbert93} suggested that the high excitation \hii\ region M17
was the best object to measure \he4\ directly and they also determined 
$\y4 \sim 0.1$.  There is mounting evidence, however, that \y4\ is
less than 0.1 in the Galaxy.  More recent calibration of the \he4\
Solar abundances yield \y4\ values lower by as much as 10\%
\citep{Grevesse96, Grevesse98, Basu04, Asplund05}.  Recent estimates
of \y4\ in M17 that include temperature fluctuations reduce \y4\ by
about 5\%, consistent with improved RRL data \citep{Peimbert02,
Quireza06}.  Furthermore, RRL observations of the lower metallicity
Galactic \hii\ region S206, which is also expected to contain little
or no neutral helium, give \y4 = 0.085 \citep{Balser06}.

Nevertheless, we adopt $\y4 \equiv y_{\rm 4GAL} = 0.1$ in equation (1)
for our ionization correction factor analysis.  We judge that for our
simple sources the ionization correction should be good to about 10\%.
At this level of accuracy all of the additional ionization effects 
considered by \he4\ abundance analyses \citep[e.g.,][]{Izotov07} 
can be neglected.  

\section{SUMMARY}\label{sec:summary}

The significance of our results for both primordial \he3\ and chemical
evolution of \he3\ lies in the fact that the abundances we find are
quite low. Hence we have paid particular attention to factors that
might lead us to measure a lower abundance than that which is actually
present.  In this paper we conclude that it is unlikely that large
ionization corrections are required for our high excitation state
\he3\ nebulae.  We use the \yp4\ abundance to select for high
excitation \hii\ regions.  We determined an ionization correction
factor for the Paper I nebular sample. There would have to be a
conspiracy between the density and ionization structure for these
ionization corrections to be underestimated by a substantial amount.
Our synthetic nebular models (\S\ref{sec:models}) do show that large
ionization corrections are possible.  Such models, however, require
that there be rather specific density and ionization structure
configurations.  Our simple ionization correction fails when \yp4\
changes significantly between the low density gas where \hep3\
dominates the emission and the high density gas where \hep4\
dominates.  The more complex the source the larger our potential error
in determining the ionization correction.  We find both numerically
and observationally that the simple sources give the best ionization
corrections (\S\ref{sec:kappa}).

For the sources that we classify as simple and which also have well
determined line parameters, the ionization correction given in
\S\ref{sec:kappa} should be accurate. Indeed, the major uncertainty is
likely to be the value adopted for $y_{\rm 4GAL}$.  Even for more complex
sources our simple ionization correction should be reasonable for any
nebula that does not have large fluctuations in excitation. We have
shown this to be the case for the complex nebula W43 where we have
single-dish and interferometry data for both radio recombination line
and continuum emission (\S\ref{sec:kappa}).

For a subset of our \he3\ sources, we use both 8\,GHz and 18\,GHz RRL
data to derive \yp4.  The two sets of observations probe different
regions of the nebula, yet we in general find very similar values for
\yp4.  This is consistent with the notion that for our sources
selected for high excitation there are not large ionization
corrections.  

We can use this simple ionization correction analysis because unlike
many of the light elements \her3\ abundances accurate to $\sim 10\%$
can yield important conclusions for both the primordial \he3\ and the
chemical evolution of \he3\ \citep{Wilson94}.  This is in stark
contrast to the situation for \he4\ where the \her4\ abundance needs
to be determined to an accuracy of $\lsim 1$\% if one is to draw
cosmologically significant inferences \citep[e.g.,][]{Izotov07}.
Furthermore, because the majority of our \he3\ nebulae are not
optically visible it is not likely that we will ever have sufficiently
detailed information about their physical properties to make a
significally better correction.

\acknowledgments{We thank the staff of NRAO Green Bank for their help,
support, and friendship.  The \he3\ research has been sporadically
supported by the National Science Foundation (AST 97-31484; AST
00-98047; AST 00-98449).  This research has made use of NASA's
Astrophysics Data System and the SIMBAD database, operated at CDS,
Strasbourg, France.}


\clearpage

%
%
\begin{deluxetable}{llrrrcrcc}
\tablecolumns{9}
\tablecaption{K-band (18\ghz) Radio Recombination Line Parmeters
\label{tab:data}}
\tablehead{
\colhead{}  & \colhead{} &
\colhead{$T_{\rm L}$}  & \colhead{$\sigma(T_{\rm L})$} &
\colhead{$\Delta v$}  & \colhead{$\sigma(\Delta v)$} &
\colhead{$t_{\rm intg}$}  & \colhead{$RMS$} & Quality \\
\colhead{Source}  & \colhead{Transition} &
\colhead{(mK)}  & \colhead{(mK)} &
\colhead{(${\rm km}\,{\rm s}^{-1}$)}  & \colhead{(${\rm km}\,{\rm s}^{-1}$)} &
\colhead{(hr)}  & \colhead{(mK)} & Factor
}
\startdata

W3.....
 & \halk\   & 579.63 &    0.76 &   27.44 &    0.04 &   11.4 &    0.97 & A \\ 
 & \healk\  &  60.85 &    0.88 &   20.94 &    0.36 &   11.4 &    0.97 & B \\ 
 & \hbetk\  & 160.73 &    0.31 &   27.80 &    0.06 &   13.0 &    0.61 & A \\ 
 & \hebetk\ &  17.30 &    0.36 &   21.21 &    0.53 &   13.0 &    0.61 & B \\ 

G133.8.....                                 
 & \halk\   &  64.99 &    0.47 &   30.06 &    0.25 &   23.6 &    0.48 & B \\
 & \healk\  &   6.04 &    0.55 &   22.94 &    2.56 &   23.6 &    0.48 & C \\ 
 & \hbetk\  &  19.63 &    0.24 &   30.14 &    0.42 &   22.2 &    0.51 & C \\ 
 & \hebetk\ &   1.41 &    0.36 &   13.01 &    4.05 &   22.2 &    0.51 & D \\ 

S206.....
 & \halk\   &  29.24 &    0.15 &   27.32 &    0.16 &   26.0 &    0.50 & B \\ 
 & \healk\  &   3.19 &    0.16 &   21.89 &    1.33 &   26.0 &    0.50 & C \\ 
 & \hbetk\  &   8.29 &    0.10 &   26.61 &    0.38 &   26.0 &    0.43 & C \\ 

S209.....
 & \halk\   &  22.24 &    0.15 &   29.70 &    0.23 &   30.8 &    0.39 & B \\ 
 & \healk\  &   1.74 &    0.17 &   23.07 &    2.71 &   30.8 &    0.39 & D \\ 
 & \hbetk\  &   7.20 &    0.14 &   29.22 &    0.67 &   30.4 &    0.41 & C \\ 

M17N.....
 & \halk\   & 387.81 &    3.28 &   33.11 &    0.32 &    8.0 &    1.97 & A \\ 
 & \healk\  &  39.74 &    3.60 &   27.69 &    2.94 &    8.0 &    1.97 & B \\ 
 & \hbetk\  & 111.53 &    1.04 &   33.27 &    0.36 &    8.0 &    0.89 & A \\ 
 & \hebetk\ &  11.93 &    1.21 &   25.32 &    3.16 &    8.0 &    0.89 & B \\ 

M17S.....
 & \halk\   & 835.20 &    5.25 &   35.29 &    0.26 &    4.8 &    2.16 & A \\ 
 & \healk\  &  77.02 &    5.41 &   34.25 &    2.98 &    4.8 &    2.16 & B \\ 
 & \hbetk\  & 245.60 &    1.46 &   35.91 &    0.25 &    4.8 &    1.57 & A \\ 
 & \hebetk\ &  22.27 &    1.51 &   34.97 &    2.96 &    4.8 &    1.57 & B \\ 

W43.....
 & \halk\   & 241.82 &    0.45 &   33.34 &    0.07 &   21.0 &    0.94 & A \\ 
 & \healk\  &  19.37 &    0.47 &   31.37 &    0.98 &   21.0 &    0.94 & B \\ 
 & \hbetk\  &  71.83 &    0.19 &   32.55 &    0.10 &   21.0 &    0.73 & A \\ 
 & \hebetk\ &   6.34 &    0.21 &   26.81 &    1.09 &   21.0 &    0.73 & C \\ 

W49.....
 & \halk\   & 417.38 &    0.60 &   29.69 &    0.05 &   20.2 &    0.85 & A \\ 
 & \healk\  &  41.31 &    0.69 &   23.94 &    0.65 &   20.2 &    0.85 & B \\ 
 & \hbetk\  & 107.21 &    0.28 &   29.55 &    0.09 &   20.2 &    0.76 & A \\ 
 & \hebetk\ &  10.67 &    0.34 &   21.55 &    1.29 &   20.2 &    0.76 & B \\ 

W51.....
 & \halk\   & 971.22 &    0.58 &   29.53 &    0.02 &    8.4 &    2.80 & A \\ 
 & \healk\  &  87.97 &    0.66 &   23.44 &    0.21 &    8.4 &    2.80 & A \\ 
 & \hbetk\  & 248.53 &    0.32 &   29.77 &    0.04 &    8.4 &    1.15 & A \\ 
 & \hebetk\ &  23.75 &    0.36 &   25.65 &    0.86 &    8.4 &    1.15 & B \\ 

NGC7538.....
 & \halk\   & 126.41 &    0.23 &   27.59 &    0.06 &   25.2 &    0.52 & A \\ 
 & \healk\  &  13.17 &    0.26 &   21.90 &    0.53 &   25.2 &    0.52 & B \\ 
 & \hbetk\  &  37.87 &    0.14 &   28.22 &    0.12 &   24.4 &    0.45 & B \\ 
 & \hebetk\ &   3.49 &    0.17 &   20.59 &    1.35 &   24.4 &    0.45 & C \\

\enddata
\end{deluxetable}

\clearpage

\begin{deluxetable}{lcccccccccc}
\tablecolumns{11}
\tablecaption{NEBULA Model Parameters$^{\rm a}$\label{tab:NEBULA}}
\tablehead{
\colhead{}  & 
\multicolumn{5}{c}{\underline{~~~~~~~~~~~~~~~Component 1~~~~~~~~~~~~~~~}} &
\multicolumn{5}{c}{\underline{~~~~~~~~~~~~~~~Component 2~~~~~~~~~~~~~~~}} \\
\colhead{Model}  & \colhead{$\theta_s$} & \colhead{$n_e$} & \colhead{$M_f$} & 
\colhead{\yp3}   & \colhead{\yp4}       & \colhead{$\theta_s$} & 
\colhead{$n_e$}  & \colhead{$M_f$}      & \colhead{\yp3}  & \colhead{\yp4} \\
\colhead{}       & \colhead{(\arcmin)}  & \colhead{(cm$^{-3}$)} & \colhead{} & 
\colhead{(10$^{-5}$)} & \colhead{}      & \colhead{(\arcmin)} & 
\colhead{(cm$^{-3}$)} & \colhead{}      & \colhead{(10$^{-5}$)} & \colhead{}
}
\startdata

A1 & 3.5 & 100 & 1.00 & 1.0 & 0.10 & 
     \nodata  & \nodata  & \nodata  & \nodata  & \nodata  \\
A2 & 3.5 & 100 & 1.00 & 0.7 & 0.07 & 
     \nodata  & \nodata  & \nodata  & \nodata  & \nodata  \\
B  & 1.5 & 500 & 0.30 & 1.0 & 0.10 & 3.5 & 100 & 0.70 & 1.0 & 0.10 \\
C  & 1.5 & 100 & 0.08 & 1.0 & 0.10 & 3.5 & 100 & 0.92 & 0.5 & 0.05 \\
D  & 1.5 & 100 & 0.08 & 0.5 & 0.05 & 3.5 & 100 & 0.92 & 1.0 & 0.10 \\
E  & 1.5 & 500 & 0.30 & 1.0 & 0.10 & 3.5 & 100 & 0.70 & 0.5 & 0.05 \\
F  & 1.5 & 500 & 0.30 & 0.5 & 0.05 & 3.5 & 100 & 0.70 & 1.0 & 0.10 \\
\enddata
\tablenotetext{a}{The distance to all model nebulae is 8\kpc.}
\end{deluxetable}

\clearpage 

\begin{deluxetable}{lccccccc}
\tablecolumns{8}
\tablewidth{0pt}
\tablecaption{NEBULA Model Results\label{tab:models}}
\tablehead{
\colhead{}  & 
\multicolumn{6}{c}{\underline{~~~~~~~~~~~~~~~~~~~~~~~X-band$^{\rm a}$~~~~~~~~~~~~~~~~~~~~~~~}} &
\multicolumn{1}{c}{\underline{K-band$^{\rm b}$}} \\
\colhead{Model}  & \colhead{$T_C$} & \colhead{$T_{\rm L}$} & \colhead{\yp3} & 
\colhead{\yp4} & \colhead{$\kappa_i$} & \colhead{\y3} & \colhead{\yp4} \\
\colhead{}  & \colhead{(K)} & \colhead{(mK)} & \colhead{(10$^{-5}$)} & 
\colhead{} & \colhead{} & \colhead{(10$^{-5}$)} &\colhead{}
}
\startdata

A1 & 0.93 & 1.10 & 1.00 & 0.100 & 1.00 & 1.00 & 0.100 \\
A2 & 0.93 & 0.81 & 0.72 & 0.070 & 1.43 & 1.03 & 0.070 \\
B  & 3.00 & 1.50 & 0.77 & 0.100 & 1.00 & 0.77 & 0.100 \\
C  & 0.93 & 0.64 & 0.57 & 0.055 & 1.82 & 1.04 & 0.067 \\
D  & 0.93 & 1.10 & 0.96 & 0.095 & 1.05 & 1.01 & 0.082 \\
E  & 3.00 & 1.10 & 0.53 & 0.085 & 1.18 & 0.62 & 0.096 \\
F  & 3.00 & 1.30 & 0.65 & 0.064 & 1.56 & 1.01 & 0.053 \\

\enddata
\tablenotetext{a}{$\nu = 8$\ghz; HPBW = 3$\arcmper$5.}
\tablenotetext{b}{$\nu = 18$\ghz; HPBW = 1$\arcmper$5.}
\end{deluxetable}

\clearpage

\begin{deluxetable}{lccccc}
\tablecolumns{6}
\tablecaption{\hii\ Region Excitation Properties\label{tab:excitation}}
\tablehead{
\colhead{} & \multicolumn{3}{c}{\underline{~~~~~~~~~~~~~~~~~~~~~~Radio\tablenotemark{a}~~~~~~~~~~~~~~~~~~~~~~}} &
\multicolumn{2}{c}{\underline{~~~~~~~~~~~~~~~~~Optical\tablenotemark{b}~~~~~~~~~~~~~~~~~}} \\
\colhead{Source}   & \colhead{\hepr4} & \colhead{$N_{\rm L}$} &
\colhead{Spectral} & \colhead{\hepr4} & \colhead{Spectral} \\
\colhead{}         & \colhead{}       & \colhead{ (s$^{-1}$)} &
\colhead{Type}     & \colhead{}       & \colhead{Type} }
\startdata

W3           & $0.076 \pm\ 0.0033$ & 49.28 & O6.5 (O5.5) & \nodata & \nodata \\
G133.8       & $0.082 \pm\ 0.0075$ & 48.96 & O7.5 (O6) &   \nodata & \nodata \\
S206         & $0.092 \pm\ 0.0029$ & 48.95 & O7.5 (O6.5) & 0.10 & O4--O6 \\
\nodata & \nodata &\nodata & \nodata & (8) & (7,8,13,17,20) \\
S209         & $0.083 \pm\ 0.0037$ & 49.82 & O3 (O4) & 0.084--0.12 & O9 \\
\nodata & \nodata & \nodata & \nodata & (8,25) & (3) \\
S212         & \nodata           & 48.67 & O8.5 (O7) & 0.088--0.10 & O5.5--O7 \\
\nodata & \nodata & \nodata & \nodata & (8,11,25) & (3,17,20) \\
S228         & \nodata             & 47.95 & B0.5 (O9.5) & \nodata & O8--B0 \\
\nodata & \nodata & \nodata & \nodata & \nodata & (3,17) \\
S235         & \nodata             & 47.86 & B0.5 (O9.5) & \nodata & O9.5--B0 \\
\nodata & \nodata & \nodata & \nodata & \nodata & (13,17) \\
S252         & \nodata             & 48.67 & O8.5 (O7) & 0.051 & O6--O6.5 \\
\nodata & \nodata & \nodata & \nodata & (24) & (5,12,14,15,19) \\
S311         & $0.071 \pm\ 0.0038$ & 48.93 & O8 (O6.5) & \nodata & O6--O6.5 \\
\nodata & \nodata & \nodata & \nodata & \nodata & (6,10,14,23) \\
\sgr{B2}     & $0.059 \pm\ 0.0182$ & 50.57 & $<$ O3 ($<$ O4) & \nodata & \nodata \\
G1.1         & $0.064 \pm\ 0.0066$ & 49.91 & O3 (O4) & \nodata & \nodata \\
M17S         & $0.092 \pm\ 0.0110$ & 50.46 & $<$ O3 ($<$ O4) & \nodata & O3--O6 \\
\nodata & \nodata & \nodata & \nodata & \nodata & (16) \\
M17N         & $0.094 \pm\ 0.0089$ & 49.75 & O4 (O5) & 0.10 & O3--O6 \\
\nodata & \nodata & \nodata & \nodata & (9,22) & (16) \\
W43          & $0.068 \pm\ 0.0052$ & 50.42 & $<$ O3 ($<$ O4) & \nodata & WN7 \\
\nodata & \nodata & \nodata & \nodata & \nodata & (2) \\
S76          & \nodata             & 47.76 & $>$ B0.5 (O9.5) & \nodata & \nodata \\
W49          & $0.079 \pm\ 0.0052$ & 50.87 & $<$ O3 ($<$ O4) & \nodata & \nodata \\
W51          & $0.079 \pm\ 0.0024$ & 50.47 & $<$ O3 ($<$ O4) & \nodata & \nodata \\
S90          & $0.070 \pm\ 0.0067$ & 48.87 & O8 (O6.5) & \nodata & O9.5 \\
\nodata & \nodata & \nodata & \nodata & \nodata & (13,14) \\
S156         & \nodata             & 48.91 & O8 (O6.5) & 0.081 & O6.5--O7 \\
\nodata & \nodata & \nodata & \nodata & (8) & (1,12,17) \\
\ngc{7538}   & $0.084 \pm\ 0.0034$ & 49.25 & O6.5 (O5.5) & 0.10 & O7 \\
\nodata & \nodata & \nodata & \nodata & (18) & (13,21) \\
S162         & \nodata             & 48.79 & O8 (O6.5) & \nodata & O6.5 \\
\nodata & \nodata & \nodata & \nodata & \nodata & (4) \\
\enddata

\tablenotetext{a}{Based on the Paper I X-band radio data that are 
appropriate for the \hep3\ ionization correction.  The
\hepr4\ abundance ratio is determined using the \hal\ and \heal\ 
RRLs (only quality factors A and B are considered).  The H-ionizing
luminosity, $N_{\rm L}$, is calculated from the radio continuum
emission.  The spectral type is determined from $N_{\rm L}$ and a
single star ZAMS model (Vacca et al. 1996; Panagia 1973, in
parentheses).}
\tablenotetext{b}{Based on optical data from the literature.  The
references are listed in parentheses.}
\tablerefs{
(1) Barlow et al. (1976);
(2) Blum et al. (1999);
(3) Chini \& Wink (1984);
(4) Conti \& Alschuler (1971);
(5) Conti \& Leep (1974);
(6) Crampton (1971);
(7) Crampton et al. (1978);
(8) Deharveng et al. (2000);
(9) Esteban et al. (1999);
(10) Feinstein \& V\'{a}zquez (1989);
(11) Fich \& Silkey (1991);
(12) Georgelin (1975);
(13) Georgelin et al. (1973);
(14) Goy (1980);
(15) Haikala (1995)
(16) Hanson \& Conti (1995);
(17) Hunter \& Massey (1990);
(18) Lynds \& O'Neil Jr. (1986);
(19) Miller (1968);
(20) Moffat et al. (1979); 
(21) Moreno \& Charvarr\'{i}a-K. (1986)
(22) Peimbert et al. (1992b);
(23) Persi et al. (1987);
(24) Shaver et al. (1983); 
(25) V\'{i}lchez \& Esteban (1996).
}

\end{deluxetable}

\clearpage

\clearpage

\clearpage

\clearpage

\end{document}